\documentclass[draft]{agujournal2019}
\usepackage{url} 
\usepackage{lineno}
\usepackage[inline]{trackchanges} 
\usepackage{soul}
\usepackage{lmodern}
\usepackage{lipsum}
\usepackage{amsfonts}

\usepackage{chngcntr}
\counterwithout{table}{section}
\renewcommand{\thetable}{\arabic{table}}

\usepackage{xcolor}
\usepackage[draft,inline,nomargin,index]{fixme}
\fxsetup{theme=color,mode=multiuser}
\FXRegisterAuthor{va}{vivi}{\color{blue}VA}

\draftfalse

\begin{document}

%
%


\title{WaveSim: A Wavelet-based Multi-scale Similarity Metric for Weather and Climate Fields}

%
%




\authors{Gabriele Accarino\affil{1,7}, Viviana Acquaviva\affil{2,3}, Sara Shamekh\affil{4}, Duncan Watson-Parris\affil{5}, David Lawrence\affil{6}}

\affiliation{1}{Department of Earth and Environmental Engineering, Columbia University, New York, NY, USA}
\affiliation{2}{CUNY New York City College of Technology, Brooklyn, New York, NY, USA}
\affiliation{3}{Lamont-Doherty Earth Observatory, Columbia University, Palisades, NY, USA}
\affiliation{4}{Courant Institute of Mathematical Sciences, New York University, New York, NY, USA}
\affiliation{5}{Scripps Institution of Oceanography and Halıcıoğlu Data Science Institute, University of California San Diego, CA, USA}
\affiliation{6}{National Center for Atmospheric Research (NCAR), Boulder, CO, USA}
\affiliation{7}{CMCC Foundation - Euro-Mediterranean Center on Climate Change, Lecce, Italy}





\correspondingauthor{Gabriele Accarino}{ga2673@columbia.edu}




%
%

%
%


\begin{abstract}
We introduce WaveSim, a multi-scale similarity metric for the evaluation of spatial fields in weather and climate applications. WaveSim exploits wavelet transforms to decompose input fields into scale-specific wavelet coefficients. The metric is built by multiplying three orthogonal components derived from these coefficients: Magnitude, which quantifies similarities in the energy distribution of the coefficients, i.e., the intensity of the field; Displacement, which captures spatial shift by comparing the centers of mass of normalized energy distributions; and Structure, which assesses pattern organization independent of location and amplitude. Each component yields a scale-specific similarity score ranging from 0 (no similarity) to 1 (perfect similarity), which are then combined across scales to produce an overall similarity measure. We first evaluate WaveSim using synthetic test cases, applying controlled spatial and temporal perturbations to systematically assess its sensitivity and expected behavior. 
We then demonstrate its applicability to physically relevant case studies of key modes of climate variability in Earth System Models. Traditional point-wise metrics lack a mechanism for attributing errors to physical scales or modes of dissimilarity. By operating in the wavelet domain and decomposing the signal along independent axes, WaveSim bypasses these limitations and provides an interpretable and diagnostically rich framework for assessing similarity in complex fields. Additionally, the WaveSim framework allows users to place emphasis on a specific scale or component, and lends itself to user-specific model intercomparison, model evaluation, and calibration and training of forecasting systems. We provide a PyTorch-ready implementation of WaveSim, along with all evaluation scripts, at: \url{https://github.com/gabrieleaccarino/wavesim}.
\end{abstract}

\section*{Plain Language Summary}
WaveSim is a new way to compare two spatial fields in weather and climate applications, by measuring how similar they are across multiple spatial scales. Rather than comparing maps point-wise, it separates each map into large-scale features and finer-scale details. It then compares the maps in three complementary ways: how strong the patterns are (magnitude), whether those patterns are in the same location or shifted (displacement), and whether their shapes and overall organization are similar (structure). We first test WaveSim using synthetic examples to better understand its behavior, under different controlled test cases. Then, we apply WaveSim to Earth System Model simulation data related to important climate patterns. This shows that WaveSim can detect similarities at multiple scales that traditional methods often miss, making it a useful tool for model intercomparison and evaluation, as well as calibration of forecasting systems.

%
%

%


%
%
%
%

\section{Introduction}
Accurately comparing gridded climate fields remains a significant challenge due to the inherently multi-scale and heterogeneous nature of Earth system processes \cite{ojima1992}. The primary purpose of field comparison is to quantify how ‘‘close’’ two spatial fields are, typically a model forecast or simulation output and a reference field or ‘‘truth’’, such as an analysis or observational data. This task, known as spatial verification \cite{briggs1997}, is crucial for many applications, including model evaluation and operational forecasting. Improved metrics for spatial verification would also directly benefit the growing field of data-driven emulators \cite{bi2023, lam2023, chen2023, kochkov2024} and bolster current efforts in evaluating and benchmarking weather and climate models' predictions, such as WeatherBench2 \cite{rasp2024}, ESMValTool \cite{eyring2020}, and ClimateBench \cite{watsonparris2022}.
Traditional point-wise error metrics, such as the Mean Squared Error (MSE), are widely used because of their mathematical simplicity and computational efficiency. However, these metrics are often found to be inadequate for spatial verification, as they compress spatially complex information into a single point-score, making them insensitive to spatial structures and to the multi-scale nature of the fields being compared \cite{briggs1997}. \\
These limitations are particularly evident when evaluating precipitation fields, which present unique verification challenges. Precipitation exhibits sparse and intermittent spatial patterns, with large areas having zero-precipitation, while small regions may contain intense, localized events \cite{buschow2019}. These features result in a patchy and highly intermittent spatial structure, which makes precipitation particularly difficult to evaluate using point-wise metrics. Even small displacements of the fields lead to the “double penalty” effect \cite{ebert2008, brown2011, kapp2018}: the forecast is penalized once for not predicting precipitation where it actually occurred, and again for predicting precipitation in the wrong location, even if the overall pattern closely resembles the reference one. This issue becomes even more pronounced when comparing high-resolution maps. \\
To address these limitations, a wide range of spatial verification methods have been proposed. A comprehensive review by \citeA{gilleland2009, ahijevych2009, gilleland2010, buschow2019} organizes these methods into four main categories: neighborhood, feature-based, field-deformation, and scale-separation techniques. Furthermore, \citeA{dorninger2018} adds distance measures as an additional category. Distance-based approaches assess the distance between forecast and observation fields by evaluating the geographical distances among all the grid points in the original space or on binary maps, after thresholding \cite{skok2022}. Notable examples include the Precipitation Smoothing Distance \cite{skok2022}, the Precipitation Attribution Distance \cite{skok2023} and the global version \cite{skok2025}, as well as application of the Fractions Skill Score \cite{skok2016, skok2018, skok2025a}.
Among these categories, scale-separation techniques are particularly relevant for this study. In this approach, fields are decomposed into spatial components at different scales, by applying a single band spatial filter (e.g., Fourier or wavelet transforms), and traditional verification metrics are applied to each scale component, separately \cite{briggs1997, casati2004, casati2007, jung2008, casati2010, weniger2017, kapp2018, buschow2020, buschow2021, casati2023}. \\
In addition to spatial verification methods, we argue that the task of evaluating how close two spatial fields are can be related to the task of assessing perceptual similarity, a concept widely explored in computer vision and Image Quality Assessment (IQA). In this context, metrics such as the Structural Similarity Index Measure (SSIM) \cite{wang2004} assess image similarity not by comparing individual pixel values, but by quantifying local characteristics such as luminance ({\it i.e.}, brightness), contrast, and structure. However, even when treating spatial fields as two-dimensional maps and interpreting values as pixel intensities, directly applying similarity-based metrics to weather and climate data remains non-trivial. Unlike natural images, such maps consist of floating-point values with physical units and unbounded dynamic ranges. Their statistical and structural properties differ significantly from those assumed in traditional IQA. Nevertheless, recent efforts have adapted these methods to climate data, through the development of the Data SSIM (DSSIM) metric \cite{baker2023}, which is similar to the SSIM, but can be applied directly on floating-point data with the aim of assessing data quality. \\ 

Building on scale separation techniques and insights from perceptual similarity frameworks, we introduce WaveSim, a multi-scale similarity metric designed for the evaluation of spatial fields. WaveSim leverages wavelet transforms to decompose fields into different spatial scales, enabling an explicit separation of information across resolutions via their corresponding wavelet coefficients. The metric consists of three complementary components, each derived from wavelet coefficients and capturing a distinct aspect of similarity: (i) {\it Magnitude} quantifies amplitude differences in the energy of wavelet coefficients; (ii) {\it Displacement} captures spatial shifts of coherent structures by comparing marginal energy distributions of wavelet coefficients; and (iii) {\it Structure} assesses the spatial organization of features, independent of their location or amplitude, detecting pattern-level similarities.
Each component provides a similarity score ranging from 0 (no similarity) to 1 (perfect similarity) at each individual scale. These component-wise scores are then combined scale-by-scale to obtain the overall similarity score. Unlike traditional point-scores computed in the original data space, WaveSim operates in the wavelet domain, allowing the integration of multi-scale, multi-component information. This results in a more interpretable and diagnostically informative assessment of similarity between complex spatial fields.\\
The remainder of this paper is organized as follows. Section \ref{sec:methods} introduces the discrete wavelet transform and presents the WaveSim methodology. Section \ref{sec:examples} reports results from the application of WaveSim, and other similarity metrics, to synthetic test cases, as well as to different modes of climate variability. Finally in \ref{sec:summary_conclusions}, we summarize the main WaveSim capabilities demonstrated through the test cases, and outline concluding remarks along with future developments.

\section{Methods}\label{sec:methods}
In this section, we provide a background on discrete wavelet transforms used for multi-scale decomposition and scale-separation. We also introduce the mathematical formulation of WaveSim, along with the datasets employed to evaluate it on synthetic test cases and on modes of climate variability in different Earth System Models.

\subsection{Discrete Wavelet Transform}\label{sec:DWT}
Discrete Wavelet Transform (DWT) is a multi-resolution analysis technique that represents a signal in terms of components localized in both space (or time) and scale (frequency). The transform is constructed from a \textit{mother wavelet} $\psi$, from which an entire family of wavelets is obtained by dyadic scaling (by powers of 2) and discrete translation:

\begin{equation}\label{eq:01}
\psi_{s,l}(x) = \frac{1}{\sqrt{2^{s}}} \psi \Big(\frac{x - 2^{s}l}{2^{s}}\Big),
\end{equation}

where $\textit{s} \in \mathbb{Z}$ is the scale index (with larger \textit{s} corresponding to coarser resolution) and $\textit{l} \in \mathbb{Z}$ is the translation index (controlling spatial or temporal shift). Each $\psi_{s,l}(x)$ is thus a rescaled and shifted version of the mother wavelet. If $\psi$ is chosen to satisfy certain admissibility and orthogonality conditions, the set $\{\psi_{s,l}(x)\}$ forms an orthonormal basis of $L^{2}(\mathbb{R})$. This means that any $f \in L^{2}(\mathbb{R})$ can be written as a sum of these basis functions: 

\begin{equation}\label{eq:02}
f(x) = \sum_{l}\sum_{s} d_{s}[l]\psi_{s,l}(x),
\end{equation}

where the wavelet (or detail) coefficients $d_{s}[l]$ at scale $2^{s}$ are obtained by projecting $f$ onto the basis.

\begin{figure}[ht!]
    \centering
    \includegraphics[width=\linewidth]{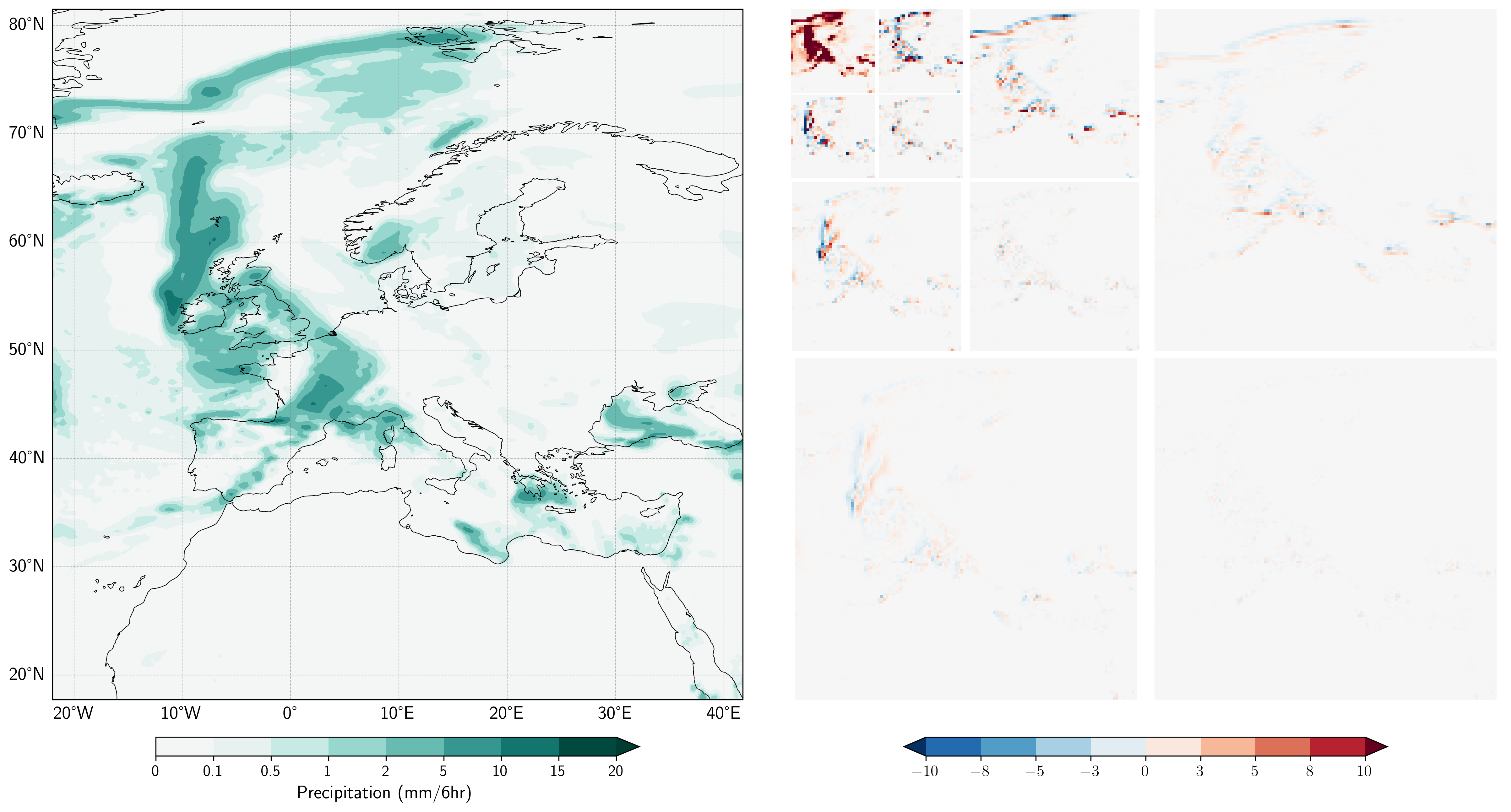} 
    \caption{Reference 6-hourly precipitation map (mm/6hr) over Europe (left panel) and the corresponding 2-dimensional 3-scales DWT using the Haar wavelet (right panel).
    The reference map, retrieved from ERA5 on 2016-01-02 at 06:00:00, was cropped to Europe ($17.5$–$81.5^{\circ}$N, $-22$–$41.75^{\circ}$E) and consists of a $256 \times 256$ grid. In the wavelet decomposition, each level yields three coefficient matrices capturing horizontal, vertical, and diagonal details, with resolution halved at successive scales ($128 \times 128$, $64 \times 64$, and $32 \times 32$ grid points). The upper-left quadrant contains the approximation coefficients, which preserve the large-scale, low-frequency structure of the field (i.e., the smoothed version of the map that is not further decomposed).
    }
    \label{fig:fig1}
\end{figure}

In practice, an infinite number of scales in equation (\ref{eq:02}) cannot be computed. To make the decomposition feasible, a scaling function $\phi$ is introduced to represent the smooth, low-frequency part of the signal. 

\begin{equation}\label{eq:03}
f(x) = \sum_{l}a_{S}[l]\phi_{S,l}(x) + \sum_{l}\sum_{s}^{S}d_{s}[l]\psi_{s,l}(x),
\end{equation}

where $a_{S}[l]$ are approximation coefficients at the coarsest scale $S$ and $\phi_{S,l}(x) = 2^{-S/2} \phi (2^{-S}x - l)$ is the scaled and translated \textit{scaling function}. Similarly to $\psi_{s,l}$, the $\phi_{S,l}$ form an orthogonal basis of functions that are smoother at the given scale $2^{S}$ that can be used to decompose the smooth residuals not captured by the wavelets (details) \cite{mallat1999, ha2021}. The mother wavelet should be selected based on the field’s spatial characteristics, ensuring an appropriate balance between smoothness, symmetry, and compactness to effectively capture its physical structures.
One of the most important properties of DWT is that both approximation and detail coefficients at scale $2^{s+1}$ can be computed from the approximation coefficients of the previous scale at $2^{s}$ \cite{mallat1989, meyer1992}. The DWT can be used to decompose 2-dimensional signals efficiently, such as images or floating-point valued matrices, like maps representing climate fields. First, equation (\ref{eq:03}) is applied separately along each axis (rows/columns). This operation generates four matrices, representing the approximation (for the Haar mother wavelet, this is simply the downscaled map at half the original resolution) and the detail coefficients at scale $2^{s}$ along three different orientations (horizontal, vertical and diagonal). The procedure is iteratively repeated on the matrix containing the approximation coefficients at scale $2^{s}$, until the desired resolution $2^{S}$ is reached. All the matrices containing detail coefficients at the intermediate and final scales are also stored, leading to a lossless representation. Figure \ref{fig:fig1} reports an example of a 2-dimensional 3-level DWT of a precipitation map across Europe. 

\subsection{Scale-separation}\label{sec:2.2}
At each scale, the DWT decomposes a field into approximation coefficients (coarse-scale, lower-frequency structures) and detail coefficients (finer-scale, higher-frequency variations). This decomposition follows a dyadic hierarchy of scales, where the spatial resolution of the coefficients is halved at each successive scale (e.g., $256 \times 256 \rightarrow 128 \times 128 \rightarrow 64 \times 64$, etc.). At every scale, the details are further separated into horizontal, vertical, and diagonal orientations providing a directional breakdown of the high-frequency content.
A fundamental property of the DWT is invertibility: the original field can be perfectly reconstructed from its approximation and detail coefficients using the Inverse Discrete Wavelet Transform (IDWT). In other words, the decomposition does not lose or discard information, but rather reorganizes it across scales and orientations.
The invertibility property enables a scale-separation procedure that isolates scale-specific contributions:
\begin{itemize}
    \item Select a particular orientation (horizontal, vertical, or diagonal) at a specific scale and set all other coefficient matrices to zero;
    \item Apply the IDWT using only the retained coefficients to reconstruct the spatial field corresponding to that specific orientation and scale, at the original map’s resolution;
    \item Repeat the procedure for the three orientations at the same scale and sum the resulting maps. This produces the scale-specific detail map, which integrates all directional contributions while remaining localized to that scale;
    \item Repeat this process across scales to yield one approximation map (containing the unresolved, coarsest-scale information) and one detail map per scale, all expressed in the original spatial resolution.
\end{itemize}

\noindent Following this approach, as illustrated in Figure \ref{fig:fig1}, at each scale (or level) we obtain four matrices of wavelet coefficients: three detail matrices (each corresponding to the full set of details at scales 1, 2, and 3) and one approximation matrix. The approximation matrix is further decomposed into detail and approximation matrices at the next (coarser) detail scale. This decomposition continues until the coarsest desired scale is reached. Together, these matrices provide a clear and interpretable separation of the field's structures across scales and orientations. As an illustrative example, a 3-scale DWT produces three maps of detail coefficients, each isolating information at a specific scale, along with one map of approximation coefficients representing the coarsest scale at the end of the scale-separation procedure.

\subsection{WaveSim}
WaveSim is a wavelet-based multi-scale similarity metric for 2-dimensional maps (e.g., climate fields). Rather than comparing two fields $X$ and $Y$ directly in their original space, WaveSim leverages their wavelet decompositions to separate information across scales. Similarity is then assessed on scale-isolated reconstructions, ensuring that contributions from different scales remain interpretable and independent. WaveSim decomposes the overall similarity score into three orthogonal components:

\begin{itemize}
    \item Magnitude similarity ($\mathcal{M}$): measures the similarity in energy magnitude, or brightness, across scales, regardless of where the energy is localized;
    \item Displacement similarity ($\mathcal{D}$): captures the degree of alignment of energy distributions along spatial dimensions (e.g., latitude and longitude). This component is designed to be invariant to global magnitude differences;
    \item Structural similarity ($\mathcal{S}$): measures the degree of preservation of structural patterns of wavelet coefficients, independently of magnitude or spatial positioning.
\end{itemize}

\noindent The overall WaveSim score combines these three components and aggregates across scales with tunable weights.
Formally, let $X, Y \in \mathbb{R}^{H \times W}$ be two input fields represented as maps of shape height ($H$) and width ($W$). For each scale $s$, we reconstruct the scale-specific maps of wavelet coefficients $\tilde{X}^{s}$ and  $\tilde{Y}^{s}$ via the IDWT procedure described in Section \ref{sec:2.2}. The resulting coefficient maps are similar to those shown in Figure \ref{fig:fig1} (right panel) but with the original map's shape, i.e. $H \times W$.
The WaveSim score is then defined as:

$$WaveSim(X,Y) = \sum_{s=1}^{S} w^{s} \cdot \Big(\mathcal{M}(\tilde{X}^{s}, \tilde{Y}^{s})^{\alpha} \cdot \mathcal{D}(\tilde{X}^{s}, \tilde{Y}^{s})^{\beta} \cdot \mathcal{S}(\tilde{X}^{s}, \tilde{Y}^{s})^{\gamma} \Big)$$

\noindent where $\alpha, \beta, \gamma \in [0,1]$ control the relative importance of each component (with $\alpha=\beta=\gamma=1$ corresponding to equal importance). Parameters $w^s$ are scale-dependent weights normalized such that $\sum_{s=1}^{S} w^{s} = 1$, which can be adjusted to emphasize specific scales depending on the application. The WaveSim framework can support both detail and approximation coefficient maps, but we only use detail  coefficients in this analysis to focus on scale-specific spatial features at each level. We provide the formal definitions of the components integrated in WaveSim next.

\subsubsection{Magnitude component ($\mathcal{M}$)} 
\noindent At each scale $s$, given the energy of wavelet coefficients for the two fields $\tilde{X}^s$ and $\tilde{Y}^s$:

$$E_{\tilde{X}^{s}} = (\tilde{X}^{s})^2, \qquad E_{\tilde{Y}^{s}} = (\tilde{Y}^{s})^2$$

\noindent We compute the mean energy per scale,

$$\bar{E}_{\tilde{X}^{s}} = \frac{1}{HW}\sum_{i=1}^{H}\sum_{j=1}^{W} E_{\tilde{X}^{s}}(i,j), \qquad \bar{E}_{\tilde{Y}^{s}} = \frac{1}{HW}\sum_{i=1}^{H}\sum_{j=1}^{W} E_{\tilde{Y}^{s}}(i,j)$$

\noindent and their relative difference:

$$\delta^{s} = \frac{|\bar{E}_{\tilde{X}^{s}} - \bar{E}_{\tilde{Y}^{s}}|}{\bar{E}_{\tilde{X}^{s}} + \bar{E}_{\tilde{Y}^{s}} + \varepsilon}. $$

\noindent We then define the magnitude component as:

$$\mathcal{M}(\tilde{X}^{s}, \tilde{Y}^{s}) = 1 - \delta^{s}$$

\noindent where $\mathcal{M}(\tilde{X}^{s}, \tilde{Y}^{s}) \in [0,1]$ and $\varepsilon$ is a small positive constant to ensure numerical stability.

\subsubsection{Displacement component ($\mathcal{D}$)} 

To account for displacement of energy spectra, we look at the normalized marginal distribution of energy along rows (latitude) and columns (longitude) of the wavelet coefficient maps:

$$\rho_{\tilde{X}^{s}}^{lat}(i) = \frac{\sum_{j}{E}_{\tilde{X}^{s}}(i,j)}{\sum_{i,j}{E}_{\tilde{X}^{s}}(i,j)}, \qquad \rho_{\tilde{X}^{s}}^{lon}(j) = \frac{\sum_{i}{E}_{\tilde{X}^{s}}(i,j)}{\sum_{i,j}{E}_{\tilde{X}^{s}}(i,j)}$$

\noindent (and similarly for $\tilde{Y}^{s}$). Normalization ensures that the marginal distributions remain independent of overall magnitude, which is already captured by the Magnitude component. We measure the divergence of marginal distributions between $\rho_{\tilde{X}^{s}}^{(\cdot)}$ and $\rho_{\tilde{Y}^{s}}^{(\cdot)}$ using the Jensen-Shannon Divergence (JSD) \cite{menendez1997}:

$$JSD(\rho_{\tilde{X}^{s}}^{lat}, \rho_{\tilde{Y}^{s}}^{lat}) = \frac{1}{2}KL(\rho_{\tilde{X}^{s}}^{lat} || m_{\tilde{X}^{s},\tilde{Y}^{s}}^{lat}) + \frac{1}{2}KL(\rho_{\tilde{Y}^{s}}^{lat} || m_{\tilde{X}^{s},\tilde{Y}^{s}}^{lat}), $$

\noindent where the mixture distribution $m$ for latitude is defined $m_{\tilde{X}^{s},\tilde{Y}^{s}}^{lat} = \frac{1}{2}(\rho_{\tilde{X}^{s}}^{lat} + \rho_{\tilde{Y}^{s}}^{lat})$, and $KL(\cdot)$ is the Kullback-Leibler divergence \cite{shlens2014}. The formulas are defined equivalently for longitude. Contrary to the KL divergence, the JSD is bounded in $[0,1]$, with $0$ indicating that the two distribution being compared are identical. Since in our similarity framework, $1$ indicates perfect similarity, we turn divergences into a displacement similarity score as follows:

$$\mathcal{D}(\tilde{X}^s, \tilde{Y}^s) = (1 - JSD(\rho_{\tilde{X}^{s}}^{lat}, \rho_{\tilde{Y}^{s}}^{lat})) \cdot (1 - JSD(\rho_{\tilde{X}^{s}}^{lon}, \rho_{\tilde{Y}^{s}}^{lon})). $$

\noindent This ensures that $\mathcal{D}(\tilde{X}^s, \tilde{Y}^s) \in [0,1]$, being the product of two quantities bounded in $[0,1]$.

\subsubsection{Structural component ($\mathcal{S}$)} 
To compute the structural component, we first flatten the wavelet coefficient matrices at scale $s$ into vectors and sort them. We define $\tilde{x}^s$, $\tilde{y}^s \in \mathbb{R}^{H \cdot W}$ as the sorted coefficient vectors at scale $s$, corresponding to $\tilde{X}^{s}$ and $\tilde{Y}^{s}$, respectively. This step retains structural patterns while removing positional information. We then center and normalize them, to remove the sensitivity to energy magnitude:

$$\bar{x}^s = \frac{\tilde{x}^s - \mu(\tilde{x}^s)}{\Vert \tilde{x}^s \Vert_{2} + \varepsilon}, \qquad \bar{y}^s = \frac{\tilde{y}^s - \mu(\tilde{y}^s)}{\Vert \tilde{y}^s \Vert_{2} + \varepsilon}$$

\noindent where $\varepsilon$ is a small positive constant added for numerical stability. Starting from $\bar{x}^s$ and $\bar{y}^s$, we compute their cosine similarity, which measures the orientation of the vectors and thus captures the similarity of their structural patterns irrespective of absolute magnitude:

$$cos(\bar{x}^s, \bar{y}^s) = \frac{\bar{x}^s \cdot \bar{y}^s}{\Vert \bar{x}^s \Vert \Vert \bar{y}^s \Vert} \in [-1,1],$$

\noindent and shift it to the $[0,1]$ range by $cos^{*}(\bar{x}^s, \bar{y}^s) = \frac{1}{2}(1 + cos(\bar{x}^s, \bar{y}^s))$. However, relying only on cosine similarity makes the metric overly permissive: two coefficient sets with very different magnitudes but similar sorted shapes may still appear structurally identical. To avoid this, we introduce a magnitude variation penalty that accounts for the relative consistency of variations in the sorted distributions:

$$P^s = 1 - \frac{\Vert \tilde{x}^s - \tilde{y}^s\Vert / n}{\mu(|\tilde{x}^s|) + \mu(|\tilde{y}^s|) + \varepsilon}$$

\noindent where $n$ is the vector length. This term penalizes structural similarity when the distributions differ significantly in relative amplitude, while remaining insensitive to global scaling already captured by the magnitude component. Specifically, it prevents cosine similarity from assigning high scores to patterns that align in shape but are arbitrarily stretched or compressed in amplitude. The final structural similarity at scale $s$ is defined as:

$$\mathcal{S}(\tilde{X}^s, \tilde{Y}^s) = cos^{*}(\bar{x}^s, \bar{y}^s) \cdot P^s$$

\noindent ensuring that the structural component is orthogonal to both magnitude and displacement, yet remains robust against disproportionate amplitude distortions.

\section{Example Applications}\label{sec:examples}
\subsection{Synthetic Test Cases}\label{sec:3.1}
\begin{figure}[ht!]
    \centering
    \includegraphics[width=0.9\linewidth]{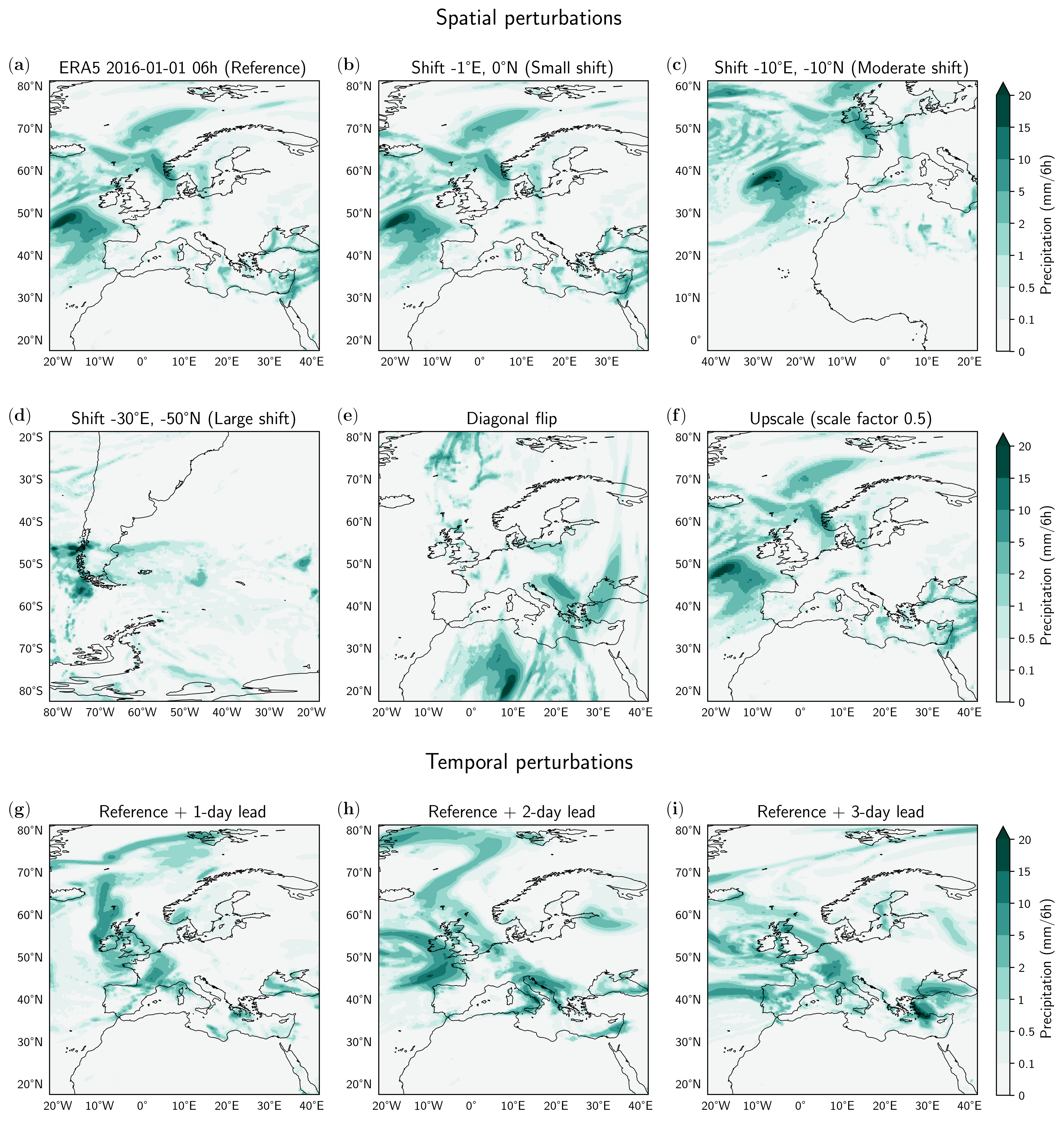}
    \caption{Synthetic test cases include both spatial and temporal perturbations (see Section \ref{sec:3.1}). Spatial perturbations are generated by applying affine transformations (Panels b–e) and a downscaling–upscaling procedure (Panel f) to a single reference map (Panel a), which depicts ERA5 precipitation patterns on January $1^{st}$, 2016. Temporal perturbations (Panels g–i) represent evolving precipitation patterns over the three consecutive days following the reference. All cases consist of maps with a resolution of $256 \times 256$ points.}
    \label{fig:fig2}
\end{figure}

To demonstrate WaveSim's behavior, we generated synthetic test cases from a single reference map of ERA5 total precipitation (mm) on January $1^{st}$, 2016, accumulated over 6 hours (Figure \ref{fig:fig2}, Panel a). From the global map, we extracted a $256 \times 256$ domain over Europe (same region as in Figure \ref{fig:fig1}, left panel) and applied a set of spatial perturbations, including: spatial shifts (Panels b–d) in which the domain is displaced by a few grid points in latitude and/or longitude, diagonal flip (Panel e) in which the domain is reflected diagonally, preserving values but altering orientation, and upscaling (Panel f) in which the map is first downscaled to half its resolution (scale factor $0.5$) and then upscaled back to the original shape, removing small-scale variability and introducing blocky artifacts. Although some perturbations (e.g., flips or more consistent shifts like in Panel d) are not representative of practical cases in spatial verification tasks, they provide controlled test cases to evaluate whether WaveSim’s components behave as expected. Beyond synthetic spatial perturbations, we also evaluate the similarity of precipitation fields at 1-, 2-, and 3-day lead times (Panels g–i) relative to the reference map (Panel a), treating them as temporally perturbed versions of the reference map. \\
Each pair of reference and perturbed fields is decomposed into four spatial scales using the Daubechies ({\it db}) mother wavelet, from which we derive WaveSim’s components and the overall similarity score. The Daubechies family of wavelets \cite{daubechies1992} consists of orthogonal wavelets with a specified number of vanishing moments, which determine their ability to represent smooth polynomial trends efficiently. A wavelet with 
{\it n} vanishing moments removes or ignores all polynomial trends up degree {\it n-1}, while capturing higher-frequency variations or details in the underlying field. Additionally, {\it db} wavelets of different orders have different {\it support}, i.e., the interval over which the wavelet  functions are nonzero. Higher order wavelets have larger supports and are more suitable to describe large fields, while lower order wavelets have more compact support. For the high resolution maps used for synthetic test cases, we use the {\it db} mother wavelet with {\it n=4} vanishing moments ({\it db4}) to accommodate the need for a larger support and the underlying rapidly varying field (better described by higher-degree polynomials, of order 4 and above).
For this evaluation, all components are given equal importance ($\alpha=\beta=\gamma=1$), and scale weights are set uniformly to $w_s = 0.25$, ensuring that their sum across the four scales equals $1$. 

We compare WaveSim with two alternative metrics: a modified version of the RMSE inspired by the skill score proposed in \cite{casati2004}, which we denote as Normalized Root Mean Squared Error (NRMSE), and the Data Structural Similarity Index (DSSIM) as defined in \cite{wang2004, baker2023}. 
 
 To build the NRMSE, we normalize the upper unbounded RMSE values and map them into a similarity scale between $0$ and $1$ as follows:

$$NRMSE(X, Y) = \exp{\left[-\left(\frac{RMSE(X,Y)}{RMSE(X,X^{*})}\right)\right]},$$

\noindent where $RMSE(X,X^{*})$ denotes the RMSE between the reference map $X$ and a randomly shuffled version $X^{*}$, representing a map generated by an unskilled model.

The DSSIM is derived from a weighted combination of three perceptual components: luminance ($l$), contrast ($c$), and structure ($s$), with weights $\alpha, \beta, \gamma > 0$:

$$SSIM(X,Y) = [l(X,Y)]^{\alpha} \cdot [c(X,Y)]^{\beta} \cdot [s(X,Y)]^{\gamma}.$$

\noindent Conceptually, WaveSim draws inspiration from the SSIM in the way multiple components are integrated and weighted, although it operates across multiple scales in the wavelet space and captures additional orthogonal information through the displacement component. Both metrics return similarity scores in the $[0,1]$ range, with $1$ indicating perfect similarity, making them directly comparable to WaveSim. 


\noindent Results on the synthetic test cases are reported in Table \ref{table:table1}. As expected, spatial shifts across latitude and longitude alter the location of precipitation patterns in the perturbed maps. As the precipitation features in the perturbed map are shifted compared to the reference, each WaveSim component is affected to some extent. This is evident in case (b), where a small longitudinal shift of $-1 ^{\circ}E$ (no shift in latitude) yields high similarity across components, with magnitude and structure remaining almost unchanged but displacement registering lower scores (ranging from $0.86$ to $0.93$). The overall WaveSim similarity of $0.87$ is consistent with the small spatial shift introduced in the perturbed map. A NRMSE of $0.71$ also reflects a moderate-to-high similarity, whereas DSSIM assigns a disproportionately low score ($0.49$) to this perturbation, suggesting that it is overly sensitive to even small geographical displacements. Case (c) considers a moderate spatial shift of $-10^{\circ}E$, $-10^{\circ}N$, and the decrease in similarity scores becomes more evident. Magnitude similarity remains high (above $0.8$); this is expected, as most of the energy in the system is still preserved in this moderate shift. On the contrary, displacement values drop significantly (down to $0.28$ at the finest scale), consistently capturing the displacement of the precipitation fronts across latitude and longitude. Structure similarity stays above $0.92$, confirming that similar structures patterns are recognized in the perturbed map, despite being spatially displaced. WaveSim's displacement component brings down the overall score ($0.28$) in a similar manner to DSSIM ($0.22$), whereas NRMSE ($0.37$) suggests a slightly higher similarity. Case (d) represents an extreme scenario in which the perturbation induces a very large shift, targeting a geographical area with no overlap with the reference map. In this case, WaveSim’s displacement similarities drop sharply across all scales (below $0.4$), indicating substantial displacement in the wavelet spectra across latitude and longitude and a lack of spatial correspondence. The magnitude component shows a mixed response, with low similarities at coarse and coarse-to-medium scales ($0.18$ and $0.32$), but higher values at medium-to-large scales ($0.97$ and $0.75$). The structural similarity remains moderately high (above $0.7$), suggesting that, despite representing distinct regions, the two fields share comparable internal organization of wavelet coefficients. Nonetheless, the low similarity values found in the magnitude and displacement components drive WaveSim to a similarity score of $0.15$, consistent with the stark mismatch of the two maps. DSSIM also reports a low similarity ($0.19$) between the maps, whereas NRMSE yields a moderately higher value ($0.43$).

Unlike spatial shifts, perturbations that flip precipitation patterns diagonally, as in case (e), are expected to preserve both structure and magnitude but substantially alter displacement. This occurs because flipping redistributes the marginal energy distributions of wavelet coefficients across locations, while keeping their amplitude unchanged. WaveSim captures this behavior distinctly: magnitude and structure components are assigned perfect similarity values across all scales (i.e., $1.0$), whereas displacement similarity collapses, especially at the finest scale ($0.14$). The overall WaveSim low score of $0.18$ derives, by construction, from the multiplicative combination of the three similarity scores, but WaveSim is able to correctly attribute the dissimilarity to its underlying source (displacement). The DSSIM reports an even lower similarity score for this perturbation ($0.09$), while the NRMSE yields an higher score ($0.35$). Both metrics fail to account for the preserved structure and magnitude. Case (f) corresponds to a perturbation involving resolution coarsening (downscaling), followed by interpolation back to the original size (upscaling). This procedure introduces blocky interpolation artifacts. As expected, WaveSim components reveal that magnitude similarity increases with scale (from $0.43$ at the finest scale to nearly $1.0$ at the coarsest), since coarse-scale variability is mostly preserved during the process. Displacement follows a similar trend, starting at $0.81$ and quickly approaching a near perfect similarity at coarse scale ($0.99$). Structure exhibits a comparable pattern, with lower score at the finest scale ($0.65$) with increasingly high similarities at the subsequent scales ($0.89-0.99$). The overall WaveSim similarity is $0.71$, capturing the degradation of small-scale patterns while recognizing the well-preserved large-scale organization. In contrast, NRMSE ($0.88$) poorly accounts for the scale-dependent differences, suggesting an higher similarity, while DSSIM ($0.79$) reports a slightly higher score than WaveSim but without the ability to pinpoint which scales are affected. This highlights WaveSim’s advantage in distinguishing the loss of fine-scale variability from the preservation of coarser features. 

Cases (g–i) represent temporal perturbations, where ERA5 precipitation maps at 1-, 2-, and 3-day lead times are compared against a common reference map (Panel a). Unlike the previously discussed use cases, these perturbations capture spatio-temporally evolving patterns and resemble the challenges typically encountered in real-world spatial verification tasks. In case (g), corresponding to the 1-day lead time, WaveSim reports an overall score of $0.38$. Magnitude and structural similarity remain high across all scales (above $0.8$ for both), while displacement drops below $0.5$ at all scales. This indicates that while the total precipitation amounts and their internal structure are largely consistent, the precipitation systems show noticeable shifts between the maps, revealing positional differences even after just one day.
At 2-day lead time (case h), the overall WaveSim similarity decreases to $0.32$. Magnitude similarity weakens, particularly at finer scales (down to $0.69$ at $\sim$100 km), while coarser scales still retain moderate agreement. Displacement remains low across scales (ranging from $0.43$ to $0.53$), reflecting increasingly misaligned precipitation systems. Structural similarity also declines slightly but remains above $0.8$, suggesting that the perturbed map retains the general organization of precipitation features, though misplaced in space.
At case (i), corresponding to 3-day lead time, the impact of spatio-temporal evolution becomes more evident. The overall WaveSim score drops to $0.25$. Magnitude similarity also shows further deterioration, particularly at fine scales ($0.59$ and $0.64$ at $\sim$100 and 50 km, respectively), and displacement registers consistently moderate-to-low scores across all scales ($0.37–0.48$), highlighting significant positional mismatches of precipitation patterns. However, structure similarity remains higher ($0.81-0.86$), though it degrades compared to 1-day and 2-day lead times. NRMSE values remain relatively high compared to both WaveSim and DSSIM ($0.43$, $0.39$, and $0.4$ for cases g–i, respectively), whereas DSSIM produces slightly lower scores ($0.26$, $0.23$, $0.24$). Overall, WaveSim appears capable of characterizing the temporal evolution of precipitation systems more effectively. Based on the results and the interpretation of its components, WaveSim not only reflects the progressive displacement of precipitation features but also captures their scale-dependent structural and magnitude changes over time.

\begin{table}[htbp!]
\centering
\caption{Similarity scores on selected synthetic test cases between the reference map (a) and perturbed maps (b-i). WaveSim components ($\mathcal{M}$: Magnitude, $\mathcal{D}$: Displacement, $\mathcal{S}$: Structure) are reported for each of the four scales, with values listed from the finest to the coarsest scale. The original spatial resolution of the field is $0.25^{\circ} \times 0.25^{\circ}$, and the scales are approximately equivalent to 50, 100, 200, and 360 km. Similarity scores are rounded to two decimal places.}
\label{table:table1}
\setcounter{table}{1}
\begin{tabular}{c c c c c c}
\hline
Panel & Test Case & NRMSE & DSSIM & \textbf{WaveSim (ours)} & WaveSim Components \\
\hline

(b) & Small shift & 0.71 & 0.49 & 0.87 &
\begin{tabular}{@{}l l@{}}
$\mathcal{M}$ & [0.96, 0.99, 0.97, 0.99] \\
$\mathcal{D}$ & [0.86, 0.92, 0.89, 0.93] \\
$\mathcal{S}$ & [0.99, 0.99, 0.98, 0.98]
\end{tabular} \\
\hline

(c) & Moderate shift & 0.37 & 0.22 & 0.28 &
\begin{tabular}{@{}l l@{}}
$\mathcal{M}$   & [0.85, 0.97, 0.88, 0.82] \\
$\mathcal{D}$   & [0.28, 0.41, 0.33, 0.30] \\
$\mathcal{S}$   & [0.95, 0.97, 0.93, 0.92]
\end{tabular} \\
\hline

(d) & Large shift & 0.43 & 0.19 & 0.15 &
\begin{tabular}{@{}l l@{}}
$\mathcal{M}$   & [0.18, 0.32, 0.97, 0.75] \\
$\mathcal{D}$   & [0.07, 0.20, 0.38, 0.35] \\
$\mathcal{S}$   & [0.71, 0.74, 0.86, 0.80]
\end{tabular} \\
\hline

(e) & Diagonal flip & 0.35 & 0.09 & 0.18 &
\begin{tabular}{@{}l l@{}}
$\mathcal{M}$   & [1.00, 1.00, 1.00, 1.00] \\
$\mathcal{D}$   & [0.14, 0.21, 0.17, 0.21] \\
$\mathcal{S}$   & [1.00, 1.00, 1.00, 1.00]
\end{tabular} \\
\hline

(f) & Upscale & 0.88 & 0.79 & 0.71 &
\begin{tabular}{@{}l l@{}}
$\mathcal{M}$   & [0.43, 0.80, 0.97, 0.99] \\
$\mathcal{D}$   & [0.81, 0.95, 0.99, 0.99] \\
$\mathcal{S}$   & [0.65, 0.89, 0.99, 0.99]
\end{tabular} \\
\hline\hline

(g) & + 1-day lead & 0.43 & 0.26 & 0.38 &
\begin{tabular}{@{}l l@{}}
$\mathcal{M}$   & [0.88, 0.82, 0.99, 0.86] \\
$\mathcal{D}$   & [0.48, 0.46, 0.42, 0.48] \\
$\mathcal{S}$   & [0.94, 0.91, 0.89, 0.96]
\end{tabular} \\
\hline

(h) & + 2-day lead & 0.39 & 0.23 & 0.32 &
\begin{tabular}{@{}l l@{}}
$\mathcal{M}$   & [0.76, 0.69, 0.83, 0.76] \\
$\mathcal{D}$   & [0.43, 0.52, 0.45, 0.53] \\
$\mathcal{S}$   & [0.92, 0.89, 0.87, 0.82]
\end{tabular} \\
\hline

(i) & + 3-day lead & 0.40 & 0.24 & 0.25 &
\begin{tabular}{@{}l l@{}}
$\mathcal{M}$   & [0.64, 0.59, 0.74, 0.80] \\
$\mathcal{D}$   & [0.37, 0.46, 0.45, 0.48] \\
$\mathcal{S}$   & [0.82, 0.81, 0.83, 0.86]
\end{tabular} \\
\hline

\end{tabular}
\end{table}

\subsection{Modes of Climate Variability}
Beyond synthetic test cases, we further evaluate WaveSim on physically relevant case studies of key modes of climate variability in Earth System Models (ESMs), including the Pacific North-Atlantic (PNA) bias and El Niño Southern Oscillation (ENSO), estimated through El Niño minus La Niña composites. The ESMs we consider belong to the Coupled Model Intercomparison Project (CMIP), the flagship international program that aims to improve climate models by comparing their outputs on a fair footing. Specifically, we select a few models from the latest phase 6, CMIP6 \cite{eyring2016, tebaldi2021} and gathered the corresponding data from LEAP-Pangeo (\url{https://leap-stc.github.io/intro.html}). Evaluation across different ESMs modes of climate variability showcase potential applications of WaveSim in assessing similarities and highlighting differences across spatial scales, which remain inaccessible to traditional evaluation metrics. In this work, we do not intend to rank ESMs or determine which one is closest to a prescribed ground truth. Instead, the focus is on interpreting point scores through different analytical components, thereby gaining insights into the underlying sources of similarity or divergence.

\subsubsection{Pacific North-Atlantic bias}
\begin{figure}[ht!]
    \centering
    \includegraphics[width=1.0\linewidth]{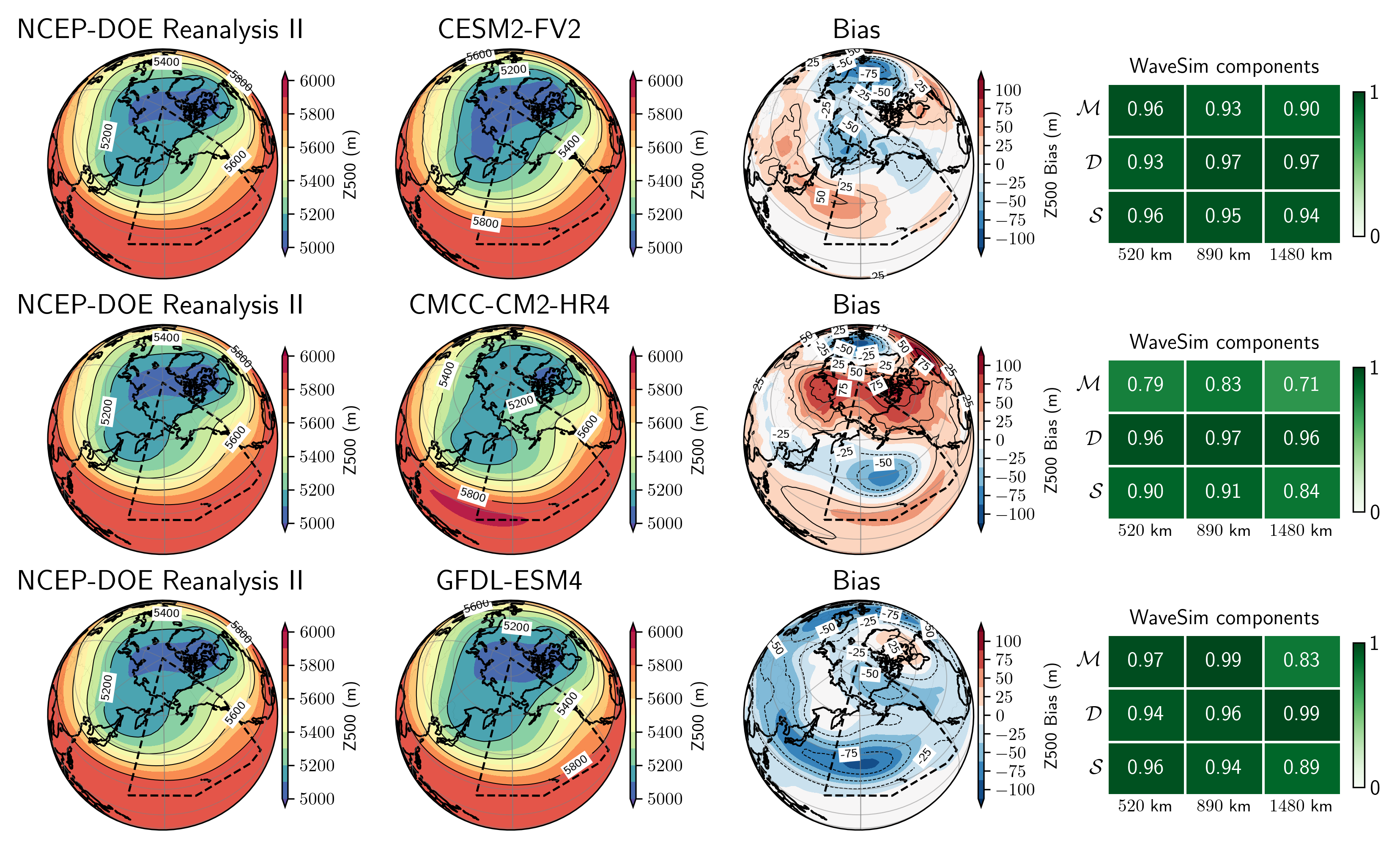}
    \caption{Comparison of DJF-averaged Z500 maps from three CMIP6 ESMs, averaged over the 1979–2014 period, to NCEP-DOE Reanalysis II. The comparison refers to the dashed region in the figure. 
    The bias panel (second from the right) shows the difference between the ESMs' averaged maps (second panel from the left) and the corresponding reanalysis (left panel), providing a visual assessment of historical biases. Z500 values are shown as contour lines in meters, using an orthographic projection. The right panel displays a heatmap of WaveSim components ($\mathcal{M}$: Magnitude, $\mathcal{D}$: Displacement, and $\mathcal{S}$: Structure) across the selected spatial scales, allowing a visual inspection of similarity values across the different scales. The corresponding similarity scores are also listed in Table~\ref{table:table2}.}
    \label{fig:fig3}
\end{figure}

We apply WaveSim to compare December–January–February (DJF) averaged (1979–2015) Z500 maps from three CMIP6 ESMs, CESM2-FV2 \cite{danabasoglu2019}, CMCC-CM2-HR4 \cite{scoccimarro2017}, and GFDL-ESM4 \cite{krasting2018}, to NCEP-DOE Reanalysis II \cite{kanamitsu2002} over the black dashed region in Figure~\ref{fig:fig3}. For consistency, ESMs Z500 data are remapped to the same grid of the NCEP-DOE Reanalysis II ($73 \times 144$ points). The selected domain of analysis, a $32 \times 32$ dyadic grid capturing part of the PNA ($161$-$240^{\circ}E$, $12.5$–$90^{\circ}N$), is decomposed into three spatial scales of approximately 520, 890, and 1,480 km, using the Daubechies-2 ({\it db2}) mother wavelet. We use {\it db2} due to its compact support, which is advantageous given the relatively small domain of application, where larger filters could exacerbate boundary effects. Additionally, {\it db2} offers a suitable balance for representing smooth fields such as Z500. Table~\ref{table:table2} reports the numerical results, including the WaveSim point score and component similarities, along with NRMSE and DSSIM for comparison. 
The assessment reveals that WaveSim is particularly sensitive to the larger bias in CMCC-CM2-HR4 (second row), yielding a point score of $0.66$. This indicates a moderate-to-high similarity, but notably lower than that assessed by the DSSIM ($0.86$) and NRMSE ($0.89$). Importantly, this difference aligns with the perceptual inspection of the bias map (ESM - reanalysis), which shows a central underestimation and a localized overestimation in the northern corner of the domain. 
Decomposing the WaveSim score into its components provides further insights. The relatively lower similarities are mainly driven by magnitude ($0.71-0.83$ across scales) and, in minor amount, by structure ($0.84-0.91$), suggesting that the energy and spatial organization of wavelet coefficients differ, albeit moderately, across multiple scales. On the other hand, displacement scores are consistently high across all scales ($0.96-0.97$), indicating that marginal distributions of energy along latitude and longitude are not significantly displaced, and the main issue lies in the amplitude and organization of patterns rather than their spatial positioning. \\
For CESM2-FV2 (first row), the evaluation yields the highest overall similarity between DJF-averaged reanalysis and simulation across all metrics, with NRMSE ($0.94$), DSSIM ($0.95$), and WaveSim ($0.84$). Unlike for model CMCC-CM2-HR4, the magnitude similarities remain high across all scales ($0.90-0.96$), and the structural component achieves consistently strong performance ($0.94-0.96$). This indicates that CESM2-FV2 reproduces both the energy distribution and the structural organization of NCEP-DOE reanalysis with higher similarity. Indeed, the contour lines in Figure~\ref{fig:fig3} between CESM2-FV2 and reanalysis appear visually very similar, except at altitudes around $5,200-5,000$ meters (green-to-yellow regions of the maps in the left and center columns), where some discrepancies emerge. The displacement component also reports strong alignment ($0.93–0.97$). Also in this case, the large scale remains well captured. \\
Finally, the GFDL-ESM4 model (third row) shows a behavior similar to that of CESM2-FV2, with overall similarity scores (NRMSE $=0.90$, DSSIM $=0.94$, WaveSim $=0.83$) higher than CMCC-CM2-HR4 but slightly lower than CESM2-FV2. The decomposition shows that, relative to CESM2-FV2, both magnitude and structure similarities decrease at the coarsest scale ($0.83$ and $0.89$, respectively), indicating stronger large-scale magnitude biases and structural deviations at this scale. In contrast, displacement values remain consistently high ($0.94–0.99$) and comparable to those of other ESMs, underscoring that the dominant discrepancies arise from the different magnitude of the field rather than its positional misalignment.

\begin{table}[htbp!]
\centering
\caption{Similarity scores (higher is better) between DJF-averaged Z500 maps from NCEP-DOE Reanalysis II and selected CMIP6 ESMs. WaveSim component values ($\mathcal{M}$: Magnitude, $\mathcal{D}$: Displacement, $\mathcal{S}$: Structure) are reported across scales, listed from finest to coarsest (520, 890, and 1,480 km, respectively), with scores rounded to two decimal places.}
\label{table:table2}
\begin{tabular}{c c c c c}
\hline
Earth System Model & NRMSE & DSSIM & \textbf{WaveSim (ours)} & WaveSim Components \\
\hline

CESM2-FV2 & 0.94 & 0.95 & 0.84 & 
\begin{tabular}{@{}l l@{}}
$\mathcal{M}$ & [0.96, 0.93, 0.90] \\
$\mathcal{D}$ & [0.93, 0.97, 0.97] \\
$\mathcal{S}$ & [0.96, 0.95, 0.94]
\end{tabular} \\
\hline

CMCC-CM2-HR4 & 0.89 & 0.86 & 0.66 & 
\begin{tabular}{@{}l l@{}}
$\mathcal{M}$   & [0.79, 0.83, 0.71] \\
$\mathcal{D}$   & [0.96, 0.97, 0.96] \\
$\mathcal{S}$   & [0.90, 0.91, 0.84]
\end{tabular} \\
\hline

GFDL-ESM4 & 0.90 & 0.94 & 0.83 & 
\begin{tabular}{@{}l l@{}}
$\mathcal{M}$   & [0.97, 0.99, 0.83] \\
$\mathcal{D}$   & [0.94, 0.96, 0.99] \\
$\mathcal{S}$   & [0.96, 0.94, 0.89]
\end{tabular} \\
\hline

\end{tabular}
\end{table}

\subsubsection{El Niño minus La Niña composite precipitation anomaly}
As a final use case, we employ WaveSim and traditional metrics to evaluate similarities across composite precipitation anomaly maps of El Niño minus La Niña events. We aim to test the flexibility of WaveSim in handling diverse contexts and scenarios, including situations where we are evaluating anomaly patterns of opposing signs rather than raw fields. \\
Specifically, we focus on El Niño minus La Niña events derived from different ESMs and compare them against a reference over the historical period $1979–2014$. For the reference, monthly Global Precipitation Climatology Project (GPCP) precipitation analyses Version 2.3 \cite{adler2016} and NOAA Extended Reconstructed sea surface temperature (SST) V5 \cite{huang2017} datasets are used to construct a baseline climatology for $1979–2014$. Anomalies are then obtained by subtracting the corresponding monthly means. The same procedure is applied to CMIP6 ESM data (CESM2-FV2, CMCC-CM2-HR4, and GFDL-ESM4), where precipitation and SST anomalies are calculated relative to each model’s own climatology over the same period. Prior to this, we remap ESM data, which have native different spatial resolutions, to the same SST ($89 \times 180$) and precipitation ($72 \times 144$) grid, ensuring consistency in the comparison. \\
ENSO years are identified consistently across observations and models using the Oceanic Niño Index (ONI), defined as the three-month running mean of SST anomalies in the Niño-3.4 region ($5^{\circ}S–5^{\circ}N$, $190^{\circ}–240^{\circ}E$). Years are classified as El Niño when ONI exceeds $+0.5^{\circ}C$ and as La Niña when ONI falls below $-0.5^{\circ}C$, relative to the $1979–2014$ climatology \cite{trenberth2025}. This thresholding ensures that only strong ENSO signals are considered. El Niño and La Niña years are then used as a proxy for precipitation anomalies. 

\begin{figure}[ht!]
    \centering
    \includegraphics[width=1.0\linewidth]{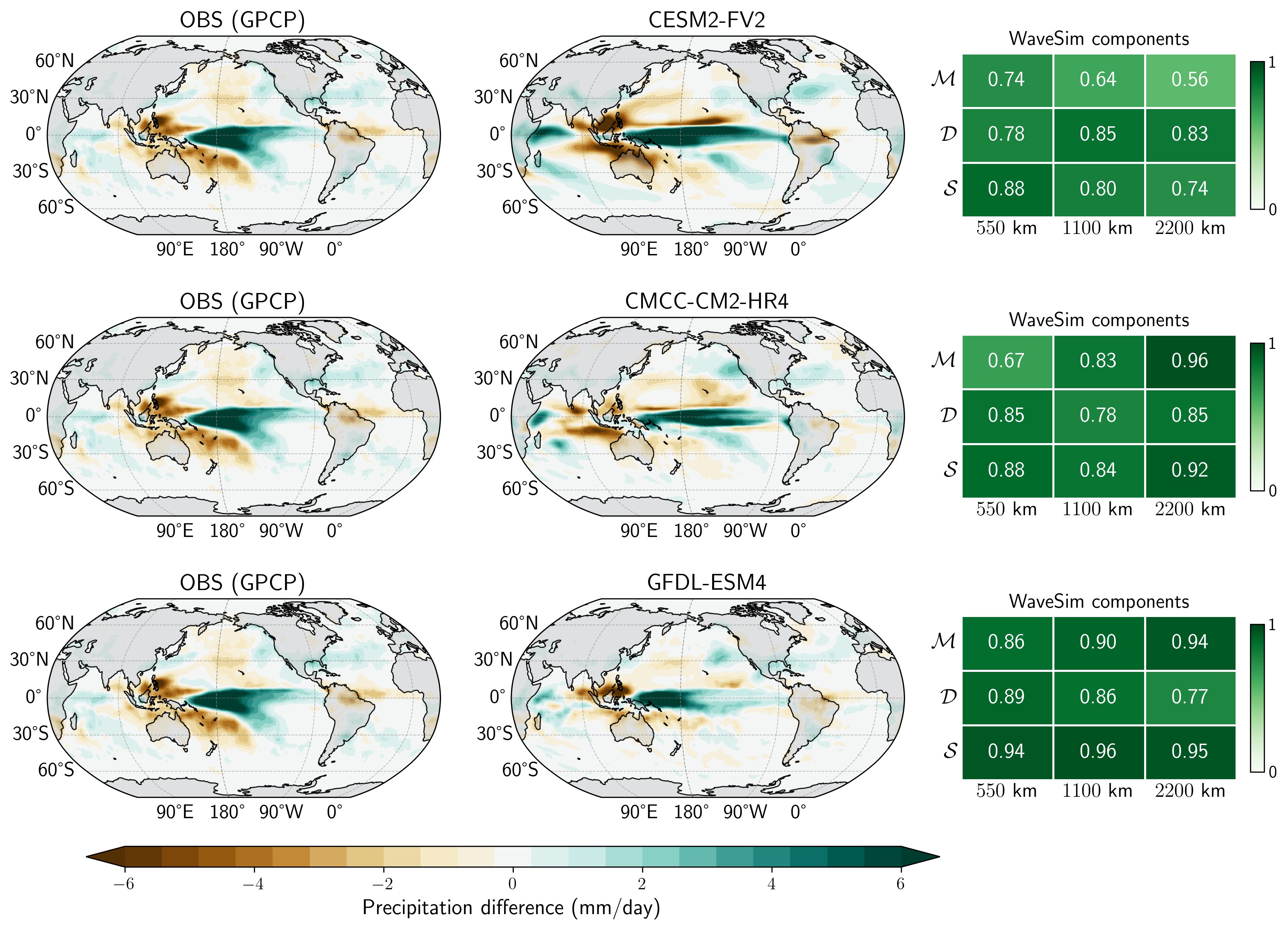}
    \caption{Composite El Niño minus La Niña precipitation anomaly during the DJF climatological season, based on the monthly climatology from 1979 to 2014, for GPCP analyses (left panel) and selected ESMs (center panel). The color bar indicates precipitation differences in mm/day. The right panel shows a heatmap of WaveSim components ($\mathcal{M}$: Magnitude, $\mathcal{D}$: Displacement, and $\mathcal{S}$: Structure) across the selected spatial scales, allowing a visual inspection of similarity values at different scales. The corresponding similarity scores are also listed in Table~\ref{table:table3}.}
    \label{fig:fig4}
\end{figure}

For both observations and models, we compute composite precipitation anomalies for El Niño and La Niña years separately over the DJF season, and subsequently derive the composite precipitation difference (El Niño minus La Niña). These composite difference maps, in Figure \ref{fig:fig4}, provide the basis for comparison with WaveSim and other metrics and answer the question of how well ESMs represent ENSO-related precipitation patterns. To calculate WaveSim, we decompose both the precipitation reference and each ESM composite into three scales using the Daubechies-2 ({\it db2}) mother wavelet, corresponding approximately to spatial resolutions of 550, 1,100 and 2,200 kilometers. The other metrics are computed directly on the composite maps without any decomposition. \\
Table \ref{table:table3} reports similarity scores between composite El Niño minus La Niña precipitation anomalies from three ESMs and GPCP analyses, which serve as the reference. Importantly, the comparison does not concern the raw precipitation fields, but rather the differences between El Niño and La Niña composites. In this setting, a high similarity score indicates that the model and the reference share comparable composite precipitation difference patterns. Across the three ESMs, NRMSE values remain around moderate values ($0.48–0.56$), suggesting that the models share similar mean amplitude of precipitation difference maps. However, this interpretation contrasts with a visual inspection of Figure \ref{fig:fig4}, which highlights perceptually distinct precipitation difference patterns, revealing that NRMSE may overestimate similarity in this context. DSSIM, by contrast, yields particularly low scores ($0.18–0.22$), indicating, at a first glance, that difference maps do not share similar structural organization. A similar behavior was already observed in the synthetic test cases: when precipitation patterns were perturbed by spatial shifts DSSIM dropped, even for relatively small displacements (e.g., Table \ref{table:table1}, rows b and c). In contrast, WaveSim provides a more nuanced view through the decomposition of similarity into magnitude, displacement, and structural components across scales. 

GFDL-ESM4 achieves the highest overall WaveSim score ($0.71$), reflecting moderate-to-high scores across all components and scales. CMCC-CM2-HR4 ($0.60$) shows weaker magnitude similarity at finer scales ($0.671$), yet strong agreement at coarser scales ($0.963$), together with high structural similarity ($0.841–0.923$) and moderately high displacement ($0.780-0.849$). CESM2-FV2 records the lowest overall score ($0.428$), largely due to magnitude mismatches at the coarsest scale ($0.555$), despite relatively high displacement ($0.777–0.854$) and structure ($0.742–0.880$).  

\begin{table}[htbp!]
\centering
\caption{Similarity scores (higher is better) obtained comparing composite El Niño minus La Niña precipitation anomaly for each ESMs and GPCP analyses in the historical period. Anomalies are relative to $1979-2014$ and averaged across the DJF season. WaveSim component values ($\mathcal{M}$: Magnitude, $\mathcal{D}$: Displacement, $\mathcal{S}$: Structure) are reported across scales, listed from finest to coarsest ($\sim$ 550, 1,100, and 2,200 km, respectively), with scores rounded to two decimal places.}
\label{table:table3}
\begin{tabular}{c c c c c}
\hline
Earth System Model & NRMSE & DSSIM & \textbf{WaveSim (ours)} & WaveSim Components \\
\hline

CESM2-FV2 & 0.48 & 0.22 & 0.43 & 
\begin{tabular}{@{}l l@{}}
$\mathcal{M}$   & [0.74, 0.64, 0.56] \\
$\mathcal{D}$   & [0.78, 0.85, 0.83] \\
$\mathcal{S}$   & [0.88, 0.80, 0.74]
\end{tabular} \\
\hline

CMCC-CM2-HR4 & 0.55 & 0.18 & 0.60 & 
\begin{tabular}{@{}l l@{}}
$\mathcal{M}$   & [0.67, 0.83, 0.96] \\
$\mathcal{D}$   & [0.85, 0.78, 0.85] \\
$\mathcal{S}$   & [0.88, 0.84, 0.92]
\end{tabular} \\
\hline

GFDL-ESM4 & 0.56 & 0.21 & 0.71 & 
\begin{tabular}{@{}l l@{}}
$\mathcal{M}$   & [0.86, 0.90, 0.94] \\
$\mathcal{D}$   & [0.89, 0.86, 0.77] \\
$\mathcal{S}$   & [0.94, 0.96, 0.95]
\end{tabular} \\
\hline

\end{tabular}
\end{table}

\newpage
\section{Conclusions}\label{sec:summary_conclusions}
In this study we have introduced WaveSim, a flexible and interpretable multi-scale similarity metric designed for comparing weather and climate maps.

Unlike traditional point-wise error metric such as MSE, WaveSim operates in the wavelet space and it is not based on pixel-level comparisons. In WaveSim, the maps are decomposed across a desired number of spatial scales through discrete wavelet transform. At each scale, three orthogonal components of similarity are evaluated and the overall WaveSim score is the results of a weighted combination of such components across scales. This design provides a richer and more diagnostic assessment of similarity, enabling users to discern whether differences (hence low similarities) arise from magnitude biases, spatial displacements, or structural misalignment. 
We note in particular that because of the multiplicative nature of WaveSim, the resulting score can be fairly low even if only one component is low, so it's always prudent to look at the breakdown by components.

WaveSim is also highly flexible and customizable. Users can adjust component and scale weights to match the requirements of specific applications, and the choice of the mother wavelet can be tailored to the nature of the field and the resolution of data.

When comparing weather and climate fields, there is no absolute “ground truth” of similarity; we rely on context and domain knowledge to assess what constitutes a meaningful level of agreement. For this reason, we calibrated WaveSim using synthetic test cases. We generated a set of controlled spatial and temporal perturbations from a single reference map, representing precipitation patterns. This process allowed us to verify that the components are well separated and behave generally as expected. For example, in the case of the "Diagonal flip" shown in Table~\ref{table:table1}, only the displacement component was affected, which was expected because the magnitude and structure of the field were unchanged. Similarly, when considering increasing lead times in forecast, i.e., for perturbations g-i in the same table, all measures of similarity decreases with time, as anticipated.

When applied to real-world ESM outputs, WaveSim provided explainable insights inaccessible to the other metrics considered. 
For instance, in the PNA-bias case study, we showed that the visual discrepancies in the Z500 contour lines of CMCC-CM2-HR4 with respect to reanalysis, are due to the cold bias ($\sim$ -75 meters) in the center of the target region and partially to the warm bias ($\sim$ +75 meters) in the top part (see Figure \ref{fig:fig3}, second row). WaveSim assigned an overall similarity score of $0.66$ to these maps and revealed that major dissimilarities arose from the magnitude component, especially at the coarsest scale ($0.71$), and in minor amount from the structural component. This intepretation aligns well with the visually percepted similarity. In contrast, the NRMSE and DSSIM assigned higher scores to these maps ($0.89$ and $0.86$, respectively), without further insights into attribution.  Furthermore, the El Niño minus La Niña composite analysis showcased WaveSim's ability to directly compare anomaly maps. In this setting, high similarity between anomaly maps indicates that models share "similar" anomaly patterns.

In conclusion, WaveSim represents a significant advance in the evaluation of spatial fields for weather and climate applications. Looking forward, we believe that WaveSim can provide a framework to support the investigation of new loss functions for data-driven emulators of weather and climate models. It is well-known that neural network-based models tend to erase small-scale variability, producing increasingly blurred estimates during roll-out on timescales typically longer than one week. This degradation is mainly due to the auto-regressive setting used to generate forecasts forward in time, as well as the use of mean loss functions such as MSE that optimize for the mean of the error distribution, thereby losing fine details and high-frequency components. Such loss functions could explicitly account for information at multiple scales as well as other components like displacement, which are typically overlooked by traditional loss functions based on point-wise comparisons. 
WaveSim could be also employed in the performance assessment of data-driven and numerical climate and weather models, as well as model and data (e.g., observations or reanalysis), with attribution of similarities to specific scales and components. Furthermore, it can aid model calibration by attributing model–data mismatches to specific spatial scales and components, helping guide parameter tuning in surrogate and emulator-based frameworks \cite{watsonparris2021, elsaesser2025}. This would enable more targeted model improvement efforts and provide deeper insight into where and why models succeed or fail. We provide a publicly available, open-source implementation of WaveSim, along with all evaluation code, at: \url{https://github.com/gabrieleaccarino/wavesim}.

\section*{Open Research Section}
Coupled Model Intercomparison Project Phase 6 (CMIP6) Earth System Model simulations were accessed through the LEAP-Pangeo platform: \url{https://leap-stc.github.io/intro.html}. 

Monthly geopotential height at 500hPa (Z500) maps have been selected from the NCEP-DOE Reanalysis II dataset: \url{https://psl.noaa.gov/data/gridded/data.ncep.reanalysis2.html}.

Monthly precipitation from the Global Precipitation Climatology Project (GPCP) Version 2.3 \cite{adler2016}, along with the NOAA Extended Reconstructed Sea Surface Temperature (ERSST) Version 5 dataset \cite{huang2017}, were also retrieved from the NOAA PSL archive: \url{https://psl.noaa.gov/data/gridded/data.gpcp.html} and \url{https://psl.noaa.gov/data/gridded/data.noaa.ersst.v5.html}, respectively.

We also provide Jupyter notebooks in the GitHub repository for data acquisition and preprocessing, enabling full reproduction of the complete pipeline and all figures presented in this paper.

\acknowledgments
We acknowledge funding from NSF through the Learning the Earth with Artificial intelligence and Physics (LEAP) Science and Technology Center (STC) (Award \#2019625). 
VA acknowledges support from a PSC-CUNY Cycle 55 grant (Award \#67628), a PIVOT fellowship grant (Award \#981849), and a PIVOT Research award (Award \#12871) from the Simons Foundation. 

%
%

\bibliography{bibliography}

%
%
%
%
%

\end{document}


%
%


\title{Supporting Information for "Insert Title"}
%
%

%
%



\authors{=Authors=}


\affiliation{=number=}{=Affiliation Address=}

%
%

%

\begin{article}

%
%

\noindent\textbf{Contents of this file}
\begin{enumerate}
\item Text S1 to Sx
\item Figures S1 to Sx
\item Tables S1 to Sx
\end{enumerate}
\noindent\textbf{Additional Supporting Information (Files uploaded separately)}
\begin{enumerate}
\item Captions for Datasets S1 to Sx
\item Captions for large Tables S1 to Sx (if larger than 1 page, upload as separate excel file)
\item Captions for Movies S1 to Sx
\item Captions for Audio S1 to Sx
\end{enumerate}

\noindent\textbf{Introduction}


\noindent\textbf{Text S1.}
%


\noindent\textbf{Data Set S1.} 


\noindent\textbf{Movie S1.} 


\noindent\textbf{Audio S1.} 


%
%


%
%
\bibliography{bibliography}
%
%
%


%
%
%
%
%

%
%
\end{article}
\clearpage